# Cash or Non-Cash? Unveiling Ideators' Incentive Preferences in Crowdsourcing Contests

Christoph Riedl
D'Amore-McKim School of Business, Northeastern University, Boston MA, USA
c.riedl@northeastern.edu

Johann Füller
Faculty of Business and Management, University of Innsbruck, Innsbruck, Austria
johann.fueller@uibk.ac.at

Katja Hutter
Faculty of Business and Management, University of Innsbruck, Innsbruck, Austria
katja.hutter@uibk.ac.at

Gerard J. Tellis
Marshall School of Business, University of Southern California, Los Angeles, CA
tellis@usc.edu

## Abstract

Even though research has repeatedly shown that non-cash incentives can be effective, cash incentives are the de facto standard in crowdsourcing contests. In this multi-study research, we quantify ideators' preferences for non-cash incentives and investigate how allowing ideators to self-select their preferred incentive—offering ideators a choice between cash and non-cash incentives—affects their creative performance. We further explore whether the market context of the organization hosting the contest—social (non-profit) or monetary (for-profit)—moderates incentive preferences and their effectiveness. We find that individuals exhibit heterogeneous incentive preferences and often prefer non-cash incentives, even in for-profit contexts. Offering ideators a choice of incentives can enhance creative performance. Market context moderates the effect of incentives, such that ideators who receive non-cash incentives in for-profit contexts tend to exert less effort. We show that heterogeneity of ideators' preferences (and the ability to satisfy diverse preferences with suitably diverse incentive options) is a critical boundary condition to realizing benefits from offering ideators a choice of incentives. We provide managers with guidance to design effective incentives by improving incentive-preference fit for ideators.

*Keywords:* Crowdsourcing; contests; innovation; ideation; incentives; non-monetary rewards; pro-social incentives.

# Introduction

Crowdsourcing innovation contests broadcast an open call to the public ("crowd") inviting them to submit ideas and problem solutions. Successful crowdsourcing contests typically draw on a wide pool of ideators, including employees, users, non-users, suppliers, distributors, and professional ideators, tapping into diverse nationalities, backgrounds, and socioeconomic groups [25,69]. The use of crowdsourcing enables both for-profit organizations such as BMW, Danone, Fujitsu, Intel, Procter & Gamble or Volkswagen, and non-profit organizations such as foundations, governments, federal agencies or NGOs to access a wide variety of high-quality ideas and solutions [2].

Despite the widespread use of crowdsourcing contests, success depends on participants' effort and creativity in contributing high-quality solutions, with earlier studies exploring factors such as prize structures [46,54,64], the number of contestants [11], or entry barriers [26]. While incentives are a key contest design element in crowdsourcing, as they are theorized to affect ideators' effort and creative performance [65], it is not clear which incentives are most effective. Previous studies show that ideators engage in crowdsourcing for a variety of reasons [1,8,11,23,45] and a mismatch between incentives and participants' motives can backfire and lead to reduced effort [29].

Although we know that ideators are motivated to participate in crowdsourcing contests for a variety of reasons, why are contests run by diverse organizations from for-profit to non-profit all using cash prizes? Given the diverse motives of the heterogeneous crowd that crowdsroucing contests aspire to attract, offering everyone the same incentive hardly seems effective [29,37]. Could crowdsourcing contest designs be improved by offering ideators a choice of different incentives?

In this paper we introduce incentive choice—which we contrast from a single, fixed incentive that is "assigned" to all ideators—as a new incentive design aspect in crowdsourcing contests. Incentive choice allows ideators to self-select between cash and non-cash prizes the incentive they prefer. We theorize that incentive choice can alleviate a central challenge in incentive design and that improving the incentive-preference fit increases effort and performance (i.e., idea quality) in crowdsourcing contests [44,56].

We further theorize that market context—the for-profit or non-profit orientation of the contest organizer—may be an important *moderator* for the effectivness of incentives. In a non-profit context, participants may be happy to volunteer their time to solve societal problems for mere recognition or praise. In a for-profit context, however, they may expect cash in return for their efforts to help organizations remain financially successful and gain competitive advantage. Market context may thus affect the effectiveness of non-cash incentives. It may also affect preference for non-cash incentives in the first place. Both of these aspects in turn may inform our understanding of boundary conditions under which offering a choice of incentives can be effective.

We address the following research questions:

1. Do ideators prefer incentives other than cash when given the choice? While it is plausible to assume heterogeneous preferences among crowdsourcing participants, we seek to quantify what the most popular incentives are. (*Quantified* in Study 1)

2. Does offering a choice of incentives improve quality and effort? (*Main effect* examined in Study 2)
3. Does market context moderate the effect of incentives on quality and effort? That is, are cash and non-cash incentives equally effective across for-profit and non-profit contexts? (*Moderator* examined in Study 3)
4. What boundary conditions may constrain the main effect of incentive choice? Specifically, if the mechanism behind the effect of offering a choice is increased preference-incentive fit, the effect may depend on the degree to which ideators actually prefer different incentives. (*Boundary condition* examined in Study 4)

In order to address our research questions, we offer evidence from four consecutive empirical studies. We start by *quantifying* incentive preferences in a realistic field experiment with over 1,000 participants (Study 1). Drawing on a broad population we offer six different incentives in a non-profit context to assess ideators' interest in choosing their own incentive and measure popularity of different cash and non-cash prizes. We establish that both cash and non-cash incentives are highly desirable. While cash is the most popular individual choice (28%), when aggregating the different non-cash options together, they account over 49%, and 23% prefer to make no choice at all. Using an econometric technique to account for self-selection in incentive choice, we establish preliminary evidence for the *main effect* of incentive choice: offering a choice *per se* improves idea quality in the observational field experiment. We provide additional evidence for the *main effect* of offering a choice from a randomized lab experiment (Study 2) using the two most popular incentive choices from the field experiment (cash and a donation). We find that offering a choice improves idea quality (but not effort). Further, we show that ideators who choose the cash incentive produce lower quality ideas and that ideators who choose the non-cash incentive exert less effort. This study also offers insights into the specific form in which the incentive choice is delivered. It reveals that ideators are not merely indifferent to incentive options and some may even choose to forego incentives entirely.

As theory suggests that the effectiveness of incentive options may depend on the market context—for-profit or non-profit—offering a choice of incentives may not be equally effective in different market contexts. Indeed, we find that market context is an important *moderator* of incentive effectivness in another randomized experiment (Study 3). We show that effort is generally lower in for-profit settings and even lower when combined with non-cash prizes. Finally, we test an important *boundary condition* in our last experiment (Study 4): despite attracting diverse populations in terms of gender, home country, or economic background, some online platforms may, over time, evolve an environment in which individuals have homogeneous incentive preferences. Using a sample of ideators drawn from a population of gig workers narrowly focused on earning income, we find no interest in non-cash incentives and consequently no benefit to offering a choice of incentives. Exploring this boundary condition suggests that incentive-preference fit (and hence improved sorting) is the driving mechanism behind the positive main effect of offering a choice (as opposed to the act of choosing itself). This suggests an important practical implication for contest designers: Offering a choice may not be effective in homogeneous ideator pools without diverse preferences.

Our paper makes three contributions to the literature on crowdsourcing design and incentive theory across different market contexts. First, we establish incentive choice as an important contest design element and explain why it works by shedding light on the underlying mechanism. This expands past work on crowdsourcing design which has focused on contest design aspects like prize structures [46,54,64], number of contestants [11], and entry barriers [26]. Second, we contribute to a better understanding of incentives in crowdsourcing by showing how for-profit and non-profit market contexts moderate the effect of incentives. Our study reveals how incentives can sometimes backfire when they are misaligned with the market context in which they are used [29,65]. Third, we extend previous research that has identified a range of reasons why individuals participate in crowdsourcing [1,8,11,23,45] by quantifying the considerable degree to which this occurs and that diverse motives result in diverse incentive choices. This guides subsequent theorizing and highlights the advantages of providing alternative incentives in contest design [58].

Our findings have practical implications as they enable managers to design more effective incentives for crowdsourcing contests. Our work suggests that in addition to cash incentives, crowdsourcing organizers should offer a choice of non-cash alternatives for participants to voluntarily choose from, with cash as a clearly marked default option to cater to ideators who are indifferent.

## Theoretical Background

### Incentives, Choice, and Market Context

*Incentives* refer to rewards that motivate or encourage someone to act. They are external to the individual and embedded in a situation. Research is increasingly clear that rewards play a complementary role to intrinsic motivation and spur desired behaviors but can also be detrimental to intrinsic motivation and reduce desired outcomes [39]. The type and level of a person's motivation together with extrinsic incentives determine a person's likelihood to become active, as well as their level of activity in terms of frequency, intensity, persistence, and performance [51,67]. Personal motives determine which incentives are perceived as attractive and how they affect an individual's behavior [52]. These reasons range from intrinsic motives (such as curiosity, interest in, and enjoyment of the task) to internalized extrinsic motives (such as skill development, making friends, or supporting others), to purely extrinsic motives associated with the outcome of their engagement (such as monetary rewards) [1,8,11,23,45]. Individuals derive more satisfaction from an activity and show higher levels of engagement the more the activity fulfills their motives [61]. Therefore, it is crucial to offer incentives that align with individuals' motivations.

Incentives are generally thought to have three kinds of effects which can be captured by a utility function with three components [9]: they value extrinsic rewards (the extrinsic motivation component), enjoy doing an activity (the intrinsic motivation component), and care about their image vis-à-vis themselves or others (the reputational motivation component). This model implies that intrinsic, extrinsic, and reputational motivations are not mutually exclusive but jointly predict behavior. However, how strongly each component factors into the utility function is specific to the individual and the context in which these incentives are employed. Individuals have been shown to differ in both their preferences for the enjoyment of a task and the image component of their utility [32]. This has resulted in a robust literature on the limits of monetary incentives [28].

There are four reasons why non-cash incentives may be especially attractive compared to cash incentives [36]. First, non-cash incentives avoid the need to justify spending money. Second, they are visible to the social environment and thus may help to build an individual's image and reputation. Third, non-cash incentives may represent an independent earning class. Because this earning class is mentally kept separate from other earnings, it may be considered especially rewarding. Fourth, individuals may mentally adjust the value of non-cash incentives depending on their personal perception and emotional reaction.

*Choice* may allow a better incentive-preference match. As individuals' utility functions are heterogeneous, for some a cash prize may be considered the most rewarding and appropriate incentive while for others the same prize may be considered inappropriate and detrimental. For example, someone may have been looking for a personal gift but received money instead. As organizers of crowdsourcing contests do not know individuals' incentive preferences upfront [48,56] it may be beneficial to allow participants to choose their incentive and thus avoid the risk of an incentive-preference mismatch. Research is scarce on how offering ideators a choice of incentives—as opposed to a single, fixed incentive that is assigned equally to all ideators—influences the effect of incentive on effort and performance. So far, some early studies have explored the effect of choice between different pay schemes such as between fixed and performance-based pay [14,57] and using observational data between monetary and symbolic awards [53]. Yet no study has investigated choice between cash and non-cash prizes more broadly nor have they investigated the underlying mechanism why such incentive choices may be effective.

*Market context* may further affect the perceived utility of an incentive and thus influence effort and quality [9,67]. For example, if you help a friend to fix a computer problem or a stranger to jump-starting a car you may expect no monetary compensation and a warm thank may be perceived as appropriate gesture. However, as IT administrator or a car mechanic you may expect monetary compensation for the same activity instead of a mere thank you. Hyman and Ariely [33] demonstrated that people categorize interactions based on whether they occur in a social or monetary market context. Depending on this classification, individuals may anticipate different incentives for the same activity. Through a series of laboratory experiments, they found that the direct impact of cash, non-cash rewards, or no incentives depends on whether the context is for-profit or non-profit (referred to as the 'market relationship'). In monetary market contexts, individuals adjust their effort and time according to the compensation offered, whereas they do not necessarily expect monetary incentives for their involvement in a social market context. Another study revealed that cash had a negative effect on the willingness to donate blood, whereas a non-cash incentive, such as a voucher of equal value, did not produce the same negative effects [40]. Thus, while previous research mainly focused on the direct effects of cash and non-cash incentives on performance, there remain open questions about how market context moderates the impact of incentives on innovation performance in crowdsourcing.

**Related Work on Incentives and Creative Performance**

Incentives represent a prominent research topic in crowdsourcing. This section provides an overview of existing studies that examine how different cash and non-cash incentives affect creative performance. Appendix 1 offers a review of existing empirical studies investigating creative or innovation performance as a dependent variable. While the review highlights the strong interest in exploring incentives and their effect on innovation performance, especially lately in the crowdsourcing context [1,11,35,65] it mainly focuses on the effect of cash rewards.

Further, existing research in psychology, education, and organizations examining the effect of extrinsic incentives on creativity remains ambiguous and sometimes show contradictory results (i.e., [3,17]). While some studies have suggested that rewards undermine intrinsic motivation and thus creative performance (e.g., [4,27]), others show that rewards lead to goal-directed behavior and thus increase creative performance [17]. More recent research has indicated that both intrinsic and extrinsic rewards may boost creative outcomes once properly adjusted to participants and context [31]. Only two studies [12,16] have explored the effects of cash and non-cash incentives on creative and innovation performance. Researchers analyzed the impact of various types of rewards, i.e., money and social cause on the contribution behavior in crowdsourcing campaigns [12] and examined intrinsic and extrinsic rewards on creative performance through five different studies [16].

While we know that participants in crowdsourcing contests have diverse incentive preferences, there is only one study that investigates the offering of choice between monetary and symbolic awards which uses observational data [53]. While the study did not find a direct effect of choice on quality, it did reveal a mediation effect of choice through effort: participants in the choice option allocated more time compared to those in the no-choice option. This study shows initial evidence that choice matters, however it does not disentangle the difference between cash and non-cash choice, the distribution of preferences, use random assignment, nor does it shed light on the underlying mechanism.

Giving ideators the choice to select their preferred incentive among various options rather than assigning a specific incentive remains an unexplored research avenue to maximize the effectiveness of incentives.[1]

Our overview further shows that most empirical studies are set either in a for-profit or non-profit context but do not compare effects across market contexts within the same study, for example by considering market context as a moderator. In summary, although we have knowledge of the direct effects of incentives, limited research exists about how choice of incentives affects performance in crowdsourcing and how for-profit vs. non-profit market contexts moderate the effect of incentives.

---

[1] Examples of existing studies investigated self-selected goal-reward levels for sales employees [10]; a lab experiment allowing individuals to choose between a fixed and performance-based pay [14]; and several studies that investigated self-selection of pay schemes allowing individuals to select a competitive (tournament) scheme or piece-rates [19,57]. In all cases, the self-selected incentive choice affected the overall expected compensation (mostly depending on the individual's skill), not the type of compensation (i.e., cash vs. non-cash).

# Research Framework

This section introduces our research framework. We start from the assumption that individuals have heterogeneous incentive preferences. We derive two hypotheses about how incentive choice and market context moderate the effect of incentives on idea quality and effort in crowdsourcing contests (Figure 1).

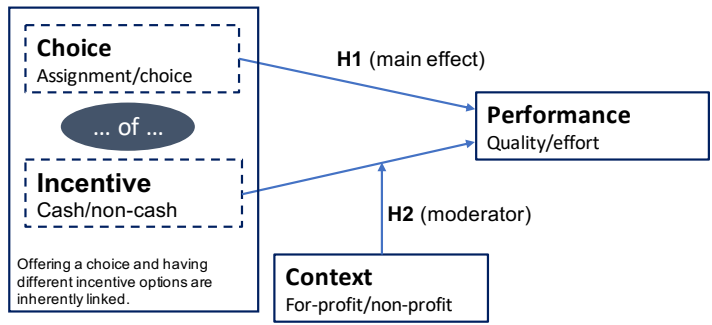

Figure 1. Research framework.

*Main Effect of Incentive Choice.* Giving ideators a choice of incentives may lead to increased effort and performance. We theorize two different pathways for a main effect of incentive choice. First, when individuals choose their preferred incentives this may lead to a better incentive-preference fit, which amplifies the effect of the incentive itself, which then spurs effort and idea quality. That is, giving ideators a choice can alleviate poor incentive-preference fit. The mechanism behind the incentive-perference fit is similar to mass customization, where customers self-configure products that meet their needs better than standardized products because they provide a better match with their preferences [22]. As incentive preferences of heterogeneous participants are not known upfront, it may make sense to let them choose their incentive [13,37,47,53]. For this mechanism to be effecitve, several conditions need to be met. Understanding these conditions will lead to a better understanding of boundary conditions under which a main effect of incentive choice may no longer materialize. First, the success of improving incentive-preference fit relies heavily on offering "appealing" choice options, which may not be an easy task. When participating in crowdsourcing contests, individuals must perceive the available options as valuable and in line with their preferences [68]. Second, in order to make choices that enhance incentive-perference fit, ideators must be aware of their own preferences, as benefits may not materialize if ideators are merely indifferent. In case of unclear incentive preference, choice could backfire by causing unnecessary effort and confusion instead of a better incentive-preference fit [50]. Finally, ideators have to have heterogeneous incentive preferences. Incentive choice will be unable to generate improved incentive-preference fit if incentive preferences are homogeneous and individuals (largely) make the same choice. Taking these boundary conditions into account, choice of incentives should lead to a better incentive-preference fit and thus increase the effectivness of incentives.

Second, an alternative reason for increased performance given a choice of incentives rests on self-determination. Giving ideators a choice of incentives could grow their sense of control over their actions, which in turn can raise intrinsic motivation [15] which increases ideators' effort and idea quality. That is, the act of choosing an incentive may in itself increase motivation by making ideators feel more self-determined and enhancing their sense of autonomy and control. We hypothesize:

*H1 (main effect): Incentive choice has a positive effect on a) quality, and b) effort.*

*Market Context as Moderator.* Research shows that the effect of incentives on effort and quality may depend on the market context and its perception as a monetary market or a social market of the crowdsourcing contest [33]. That is, market context may moderate the effect of different incentives because, depending on the context, individuals may have different incentive *expectations* for the same activity. Consider getting help for a painting job or preparing tax returns. In such situations, one may ask either a friend or a professional [33]. While a friend may help without expecting a return [7], a professional expects money for their time and effort [20]. Here, the relationship the person seeking help has with the person providing the help, moderates the effect of the (non-)cash incentive. It will work well in one context but can backfire in another.

According to *relational incentive theory* [21,29] people categorize the context in which an interaction takes place into one of four relationship categories: common sharing, authority ranking, equality matching, and market pricing. They adjust their participation behavior and reward expectations according to the classified context. In common sharing (CS), individuals consider their engagement as contributing to a common goal and supporting the group to solve a pressing problem without asking for returns because of shared beliefs, solidarity, and altruistic reasons. In authority ranking (AR), individuals accept a hierarchical order. They engage in relationships to learn from superiors, fulfill their duties as good citizens, or conform to the authorities. Equality matching (EM) is characterized by reciprocity, balance, and tit-for-tat. Individuals engage in, e.g., carpooling or dinner party invitations because they trust that others will reciprocate at a later time. Market pricing (MP) refers to situations dominated by cost-benefit calculations, where personal gain determines if one engages in an activity or not. While the first three relationship patterns (CS, AR, EM) are social in nature, the MP schemata is based on economic exchange. Thus, previous research suggests simplifying the model and subsuming the four categories into two categories, *social markets*— consisting of the CS, AR, and EM relationships, and *monetary markets* – referring to MP [33]. They investigate the main effect of different market contexts and find that students show higher intentions to help moving a sofa, spend more time in an online experiment, and show higher efforts in solving a serious of puzzles depending on the cash or non-cash incentives offered [33]. They conclude, while monetary markets are sensitive to compensation, social ones are not. In monetary markets, effort is directly related to the amount of compensation. In social markets, effort is shaped by altruism.

Based on this insight, we theorize that the same incentive will be perceived differently depending on the market context and that market context will thus moderate the effect of the incentive [29]. Specifically, we expect the effect of non-cash incentives to weaken in monetary market contexts.

While for-profit organizations develop new solutions to stay competitive and generate profits for their shareholders, non-profit organizations innovate to solve pressing problems and create common goods that benefit everyone. We therefore suggest that *for-* *profit* contests be perceived as *monetary markets*, while *non-profit* contests be perceived as *social markets*. When the incentive is in line with the market context (such as a cash incentive in a for-profit setting) this should strengthen the effect of the incentive. Conversely, when the incentive is misaligned with the market context (such as a non-cash incentive in a for-profit setting), the effect of the incentive should weaken. Further, as participants adjust their incentive preferences and behaviors according to the market context, the for-profit status of the contest organizer may activate ideators' extrinsic motives and lead to expectations of cash. Alternatively, those engaging in non-profit settings may not expect to get paid but may appreciate a non-cash reward like a gift or praise. Non-cash prizes may in fact constitute a mismatch in a for-profit contest setting, where organizers benefit economically from the crowd's contribution while the ideators receive only a non-cash prize. Ideators may expect a fair monetary compensation when organizers generate successful returns [30]. Thus, we theorize:

*H2 (moderator): Market context moderates the effect of cash and non-cash incentives on a) quality and b) effort.*

## Multi-Study Overview

We conducted four studies (see Figure 2) to *quantify* incentive preferences in a non-profit field study (Study 1), test the *main effect* of incentive choice (Study 2), investigate whether market context *moderates* the effect of incentives on quality and effort (Study 3), and finally explore preference heterogeneity as a *boundary condition* (Study 4). Together, these four studies paint a more complete picture of how incentive choice and market context collectively affect effort and idea quality in crowdsourcing contests.

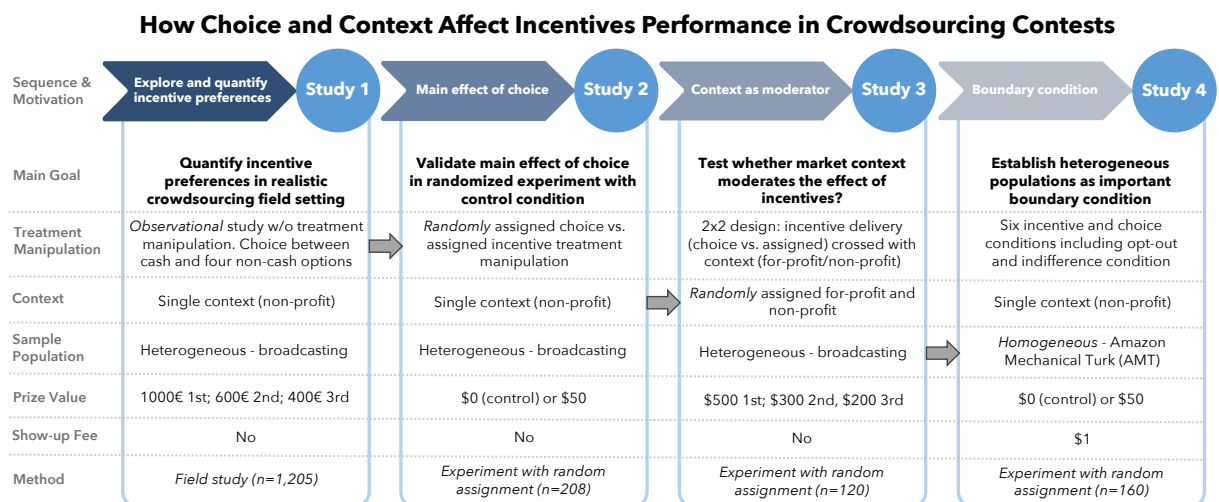

Figure 2. Summary of sequential study design.

# Study 1: Quantify Non-Cash Incentive Preferences in a Field Experiment

**Design and Empirical Setting**

To *quantify* ideators' incentive preferences and offer preliminary evidence for a *main effect* of incentive choice, we set up the "Scraplab" crowdsourcing contest. It was hosted on a leading contest platform (www.hyvecrowd.com). The contest dealt with up-cycling and the goal was to create products out of recyclable materials instead of producing waste. Since the contest contributes to the Sustainable Development Goals of the United Nations and aims to create impact rather than increasing corporate profits, the context can be classified as non-profit. This topic seemed to be appropriate as it does not require specialized knowledge, skills, or familiarity with existing brands. Further, because resource shortage affects everyone, anyone can have ideas on how to solve it.

The submission of an idea consisted of a visual design in the form of photographs, a textual description, and a list of materials used (see Appendix 2 for sample designs). As is common in crowdsourcing contests, ideators could make multiple idea submissions, create a profile page to share personal information, interact with each other by commenting on design submissions, provide feedback through ratings, and promote designs by sharing them on Facebook.

The contest was designed as a rank-order tournament [43] in which the three highest-rated ideas would receive prizes valued at 1,000€ (1st), 600€ (2nd), and 400€ (3$^{rd}$). During registration, participants were offered the option to choose among a cash prize and five non-cash prizes of the same value. The non-cash prizes included, a donation to a charity of choice (altruistic motives [34]); a short internship (same renumeration as the cash prize); career advancement motive [41]; participation in a workshop to improve their own design innovations (need for a solution [34] or a funded party with friends (hedonic motives [5,9]). As the incentive question was optional, participants who didn't select one of the five options were assigned to the default cash incentive.

*Recruiting and participants.* To reach a large and global audience, we advertised the design contest globally in various design and sustainability communities, including design schools. The competition was open for submissions for ten weeks. During the contest period, the website had 16,686 unique visitors (unique IP addresses). 1,205 participants registered and were exposed to the incentive-choice question during their upfront registration. 924 participants (77%) answered the question and chose their preferred incentive. 281 participants (23%) did not make a choice. 259 participants submitted one or more ideas/designs (587 ideas in total) and thus are labeled as ideators. Among the 259 ideators, 118 answered the optional incentive preference question while 141 did not and were automatically assigned to the default cash option. At the end of the contest, three ideators were awarded prizes for their designs. Ideators from 64 different countries participated in the contest. The highest concentration of ideators was from the United States of America (48%) and most were female (75%).

**Data Sources and Measures**

We used three main sources of data in our study: (1) registration data on ideators' demographics, occupation, and preferred incentive; (2) log-file data from the online platform running the contest to explore participants' behavior; and (3) data from an independent external consumer panel using experienced workers on Amazon Mechanical Turk to assess the quality of design submissions. Appendix 3 shows descriptive statistics and correlations of our individual-level measures.

**Independent variables and controls.** The key variable of interest is the ideators' *Incentive Preference*, which was an optional question on the registration survey. The question was not forced so that ideators could choose whether to select an incentive or not. We focused our analysis on contrasting ideators who chose the cash prize from those who chose any of the other non-cash prizes. Data on ideators' gender was from the registration survey, as well as imputed from the first names and profile pictures of ideators where necessary.[2] Gross Domestic Product (GDP) data from each ideators' current country of residence was collected using 2011 numbers from the World Bank. For ideators with missing values for country of origin (50 instances), we substituted the GDP sample mean [62].

Because the key outcome measure of interest in this study is the quality of ideas, it is critical to control for ideators' ability, which may affect the quality of their designs [42]. Ability is generally unobserved and difficult to capture reliably. As an imperfect measure, we included a measure indicating whether an ideator is a professional designer. Professional designers are expected to have relevant experience in design tasks like that of our contest, by having previously engaged in design work full-time for an extended period. Experience is probably the most widely used proxy for expertise [18]. Thus, based on information that ideators provided during the registration procedure, we included an ideator's status as a professional designer as a proxy for expertise (1=professional designer; 0=not professional designer).

Data from the online platform includes information on the time an ideator first registered, and the number of designs, comments, and ratings he/she submitted. We included controls for the number of *Submitted Ideas, Submitted Comments,* and *Submitted Ratings*, as these provide ideators with the ability to learn from both their own direct experience and from observing the work of others [60]. Lastly, we included a control for the number of ideas that were already submitted to the contest just prior to the signup of an ideator. The number of prior ideas is an easily observable signal to potential contributors of how competitive a contest is. As such, it may affect ideators' choice of incentive.

**Dependent variable.** This is the quality of each ideator's best idea. We followed standard practice in ideation studies [38,66] and measured the dependent variable, *Design Quality*, for all designs submitted to the contest. We used an outside panel that followed a relative assessment technique [4]. We recruited an independent jury who were blind to the research propositions. They were experienced workers on Amazon Mechanical Turk. We collected five evaluations for each design, resulting in 2,927 ratings of each of the six assessment items (17,562 ratings in total) from a total of 77 different raters (Appendix 2 provides details on the method, the assessment items, and some design examples). We analyzed how incentive choice affects ideators' probability to become active (Appendix 4).

---

[2] Two researchers independently coded gender; discrepancies were discussed and resolved; 12 out of 125 instances of missing gender information could not be coded and were subsequently excluded from the analysis.

## Results

**Quantifying incentive preferences.** We find that when given the choice, 77% of ideators (N=924) chose an incentive and revealed their preference (Table 1). Almost half of ideators (49%, N=591) preferred a non-cash incentive over cash. Only 28% (N=333) actively chose the cash incentive. The most popular non-cash incentive was a donation with 14% (N=171), followed by the internship with 14% (N=170), the workshop incentive 11% (N=127), and the party incentive 10% (N=123). Personal characteristics such as gender and economic background are weak proxies for non-cash incentive preferences (Appendix 5). This suggests that while overall non-cash incentives are very popular, there are many different forms such incentives could take.

Table 1. Study 1 – Descriptive statistics of incentive preferences for participants who made at least one design submission.

|            | N     |        | Percent Female | Percent > 0 Effort |
|-----------:|:-----:|:------:|:--------------:|:------------------:|
| No Choice  | 281   | (23%)  | 56%            | 50%                |
| Cash       | 333   | (28%)  | 72%            | 21%                |
| Donation   | 171   | (14%)  | 82%            | 6%                 |
| Internship | 170   | (14%)  | 80%            | 18%                |
| Workshop   | 127   | (11%)  | 97%            | 5%                 |
| Party      | 123   | (10%)  | 90%            | 1%                 |
| Total      | 1,205 |        | 75%            | 21%                |

**Main effect of incentive choice on quality.** Since Study 1 is an observational study that did not permit the randomized assignment of incentives, the analysis of the main effect of choice is complicated as the analysis needs to account for self-selection of the assignments. We analyze our data using a Tobit-5 switching regression (this is also known as the Roy model [59] to distinguish the cash vs. non-cash incentive effect from the influence of choice (see Appendix 2 for details on the Tobit-5 model). To establish whether there is main effect of offering a choice we need to establish two findings. First, we need to determine whether there is a performance difference among ideators who chose the cash vs. the non-cash incentive. Second, we need to examine whether the correlation of the error terms of the self-selection component in the model are positively correlated. A positive correlation would signify that ideators who chose the incentive performed better than they would have in a hypothetical scenario in which they were randomly assigned to that incentive. Regarding the first condition, we find systematic performance differences between ideators who chose cash and those who chose a non-cash prize. Specifically, we observe a significantly higher effect of the non-cash incentive on idea quality compared to the cash incentive (direct incentive effects in Table 2; testing for equal coefficients comparing $\beta_{non-cash} = 3.6$ to $\beta_{cash} = 3.0$; $\chi^2(12) = 10.29$; $p <$

0.001). Regarding the second condition, we find that the estimated correlation coefficients between the error term of the selection equation and the outcome equations (the ρ1 and ρ2 estimates in the table), are both large and significantly different from zero ($p < 0.001$). Since ρ1 is positive and significantly different from zero, the model suggests that individuals who chose the non-cash prize produced designs of higher quality (than a random individual from the sample). Conversely, since ρ2 is negative and significantly different from zero, the model suggests that ideators who chose a cash prize produced designs of lower quality (than a random individual from the sample would have).

Table 2. Study 1 – Selection and outcome equations comparing the choice of the cash prize over any of the non-cash prizes. Standard error in parentheses. Sample: 118 ideators who answered the incentive question and submitted at least one design.

| Selection Equation | Pr(Choose Cash) | |
|---|---|---|
| Intercept | 0.63 | |
| | (2.03) | |
| log(GDP) | −0.07 | |
| | (0.20) | |
| Western | −0.62*** | |
| | (0.20) | |
| Female | −0.30 | |
| | (0.20) | |
| Designs Prior to Registration | 0.00*** | |
| | (0.00) | |
| Outcome Equation | Design Quality | |
| | Non-Cash | Cash |
| Direct Incentive Effect (Intercept) | 3.55*** | 3.03*** |
| | (0.11) | (0.10) |
| Professional experience | −0.30 | 0.11 |
| | (0.12) | (0.10) |
| Submitted Designs | 0.10 | 0.13*** |
| | (0.06) | (0.05) |
| Submitted Comments | −0.00 | 0.05** |
| | (0.02) | (0.02) |
| Submitted Ratings | 0.008 | −0.005** |
| | (0.01) | (0.00) |
| Choice Effect | | |
| ρ1 | 0.78*** | |
| | (0.05) | |
| ρ2 | | −0.73*** |
| | | (0.06) |
| Num. obs. selection eq. | 118 | |
| Num. obs. outcome eq. | 49 | 69 |
| AIC | 530.72 | |
| Log Likelihood | -246.36 | |

***$p < 0.01$; **$p < 0.05$; *$p < 0.1$

Based on these estimates, we can predict a counterfactual of how ideators who chose the non-cash incentive would have responded if they were assigned to the cash incentive (i.e., we compute $y_i^{O1} - E(y_i^{O2}|y_i^S = 1)$) and vice versa to estimate the direct benefit of offering ideators a choice. This suggests an 11% increase in design quality from offering ideators a choice of incentives compared with simply offering cash to everyone. That is, using an econometric approach to account for self selection, the observational data from Study 1 provides preliminary evidence for a *main effect* of incentive choice on idea quality supporting H1a. Since we do not know how much time ideators spent on creating their upcyled designs, we cannot test hypothesis H1b (on effort) in Study 1.

### Summary

Participants had diverse incentive preferences and chose both cash and non-cash prizes when given the choice. While cash was the most popular choice, almost half of ideators chose one of the non-cash options, and about one quarter preferred to not make a choice at all (receiving the default cash option). We also find preliminary evidence that incentive choice had a significant *main effect* on idea quality: Offering ideators a choice of incentives is an effective strategy for contest organizers to unlock additional value independent of the incentive offered, supporting H1a.

## Study 2: Main Effect of Incentive Choice in Online Experiment with Random Assignment

To explore the preliminary evidence of a *main effect* of incentive choice, Study 2 uses random assignment to give some ideators a choice of incentives while others had no such choice. Further, as the field study left unanswered why some participants did not actively choose a prize, Study 2 also tests participants' indifference between cash and non-cash prizes as well as the decision to opt out of receiving any prize.

### Method

**Procedure.** We implemented an online experiment that mirrored the task of an ideation crowdsourcing contest (Figure 3). First, subjects completed a pre-treatment survey to answer demographic questions, as well as a practice task (a standard "unusual uses" creativity task [16], and answered three items from the Intrinsic Motivation Battery [51]; Cronbach's $\alpha = .9$). Second, we performed a treatment manipulation in which subjects were either randomly assigned a prize ("assigned" condition), were asked to choose a prize ("choice" condition), or were not offered a prize ("control" condition). Third, subjects completed the main ideation task. Fourth, we used an external panel to evaluate idea quality.

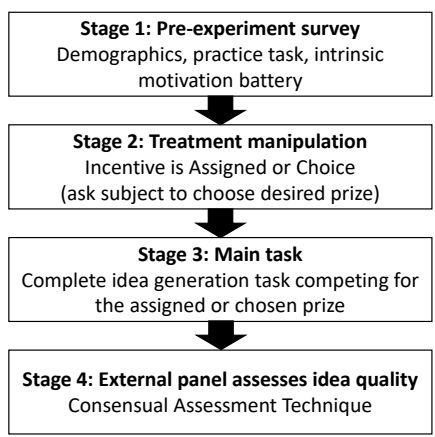

Figure 3. Study 2 – Study procedure and treatment flow.

The main ideation task asked subjects to submit ideas to reduce water consumption, which they entered through a free-form text field. Again, as in Study 1, the context can be classified as non-profit as the ideation task contributes to the Sustainable Development Goals and aims to create impact rather than increase corporate profits. In addition to the idea itself, we also measured effort in terms of time (in seconds) that participants devoted to submitting their idea.[3] Individuals could type as many ideas as they wanted (in situations where multiple ideas were submitted, we calculated the overall effort as the total number of seconds spent entering all of the ideas).

**Recruiting**. To recruit participants, we advertised the experiment on the Volunteer Science platform, an online laboratory for experiments in social psychology. We performed no specific recruiting to attract participants for our study (see [59] for details and validation of the Volunteer Science platform).

**Treatments.** All subjects competed either for no prize at all, a $50 cash prize, or a $50 donation to a charity of their choice. Participation was entirely voluntary and no additional compensation (such as a flat show-up fee) was paid. The treatment manipulation was whether the prize was randomly assigned or whether the subject got to choose what the prize was. We implemented three alternative versions of the choice treatment manipulation to explore the kinds of choice options contest designers may consider:

1. Assigned no prize (control condition)
2. Assigned a $50 cash prize (i.e., no choice offered)
3. Assigned a $50 donation to a charity of choice (i.e., no choice offered)
4. Choice between either a $50 cash prize or a $50 donation
5. Choice between either $50 cash or a $50 donation, measured on a 7-point Likert scale. We treated the central values (3, 4, 5) as indicating indifference between cash and non-cash incentives, while we treated strong (1) and weak (2) preference for cash as cash and strong (7) and weak (6) preference for non-cash as non-cash. We informed ideators that their reward would be determined randomly using the proportions of their choice (i.e., 50:50 chance between cash and non-cash if they chose the middle point "4" on the Likert scale; 43:57 chance if they chose "5" and so on).
6. Choice between either accepting a $50 cash prize or opting out of prizes entirely.

We performed (streaming) random assignment of participants to treatment conditions as is common in online experiments where the total number of participants is not known ex ante. We assigned more ideators to the choice conditions to reflect the fact that these conditions have several nested sub-conditions.

**Sample and Measures**. We stopped the experiment after 40 days, at which point 221

---
[3] We also implemented a mechanism to detect if they would simply copy & paste text into the text form (none did).

individuals had completed it and 208 ideas had been submitted (e.g., we removed ideators who submitted ideas such as "I don't know;" Appendix 6). Four individuals dropped out after having been assigned to a treatment condition. The dropout was not correlated with treatment condition ($\chi^2(5) = 4.78$). Women and men participated in equal proportion (51% female) with equal proportion in each treatment ($\chi^2(5) = 2.06$). Participants were young (94% reported age between 18-24) and mostly from the USA (92%).

We measure idea quality using Amabile's Consensual Assessment Technique from an outside panel recruited through Amazon Mechanical Turk, following the same method and procedure as in Study 1. We collected 1,440 quality ratings in total from 93 different raters who performed an average of 15.5 ratings each. Inter-coder reliability is excellent (0.836); Cronbach's alpha is good (0.84).

### Results

**Quantifying incentive preferences.** Again, we find heterogeneous incentive preferences: 56% chose the cash incentive, and 44% the donation incentive when given a choice (see Appendix 6). Individual-level characteristics like gender, age, home country, or intrinsic motivation are no significant predictors of incentive preference. The field study did not address whether ideators may simply have been indifferent to our prize options, so we also explored whether ideators would opt-out or are indifferent. When given the option to forgo a cash prize, 29% of participants chose to opt out. We find that 37% of individuals who were given a choice between cash and non-cash indicated that they were indifferent between the two (they selected one of the middle points on the Likert-scale).

**Main effect of incentive choice on quality and effort.** We used OLS regression to analyze the *main effect* of incentive choice on idea quality and a negative binomial regression for the effects on effort (Table 3). This analysis contrasts the two levels of the treatment conditions of offering a choice vs. assigning a fixed incentive to ideators. First, we find that offering ideators a choice, *per se*, improved idea quality (Model 1: $\beta = .17$; marginally significant at $p = 0.076$; supportting H1a) but not effort (Model 3: $\beta = .263$; n.s.; rejecting H1b). This suggests that incentive choice increases idea quality by 6.6%.

To better understand the main effect of incentive choice, it is crucial to understand which incentives are most effective when chosen. We explore this through the interaction between the choice treatment (choice vs. assigned) and the incentive. The interaction between choice and cash shows a significant negative effect on idea quality ($\beta = -.46$; $p < 0.05$). We also find a significant interaction between choice and the non-cash incentive on effort ($\beta = .85$; $p < 0.001$). This suggests that the effectiveness of incentives on both idea quality and effort is moderated by the form in which incentives are delivered such that choice decreases the effect of cash incentives on idea quality and increases the effect of non-cash incentives on effort.

Table 3. Study 2 – Regression analysis. Omitted category: Assigned, no-prize. Note that Prize: Indifferent implies Choice: Yes.

| Dependent Variable | Idea Quality | | Effort | |
| --- | --- | --- | --- | --- |
| | (1) | (2) | (3) | (4) |
| <u>Treatments</u> | | | | |
|   Choice: Yes | 0.17* | 0.52** | 0.263 | 0.12 |
| | (0.10) | (0.25) | (0.212) | (0.35) |
|   Prize: Cash | 0.27** | 0.53*** | 0.607*** | 0.83*** |
| | (0.14) | (0.20) | (0.194) | (0.30) |
|   Prize: Non-Cash | 0.37** | 0.55*** | 0.829*** | 0.32 |
| | (0.15) | (0.21) | (0.314) | (0.30) |
|   Prize: Indifferent | na | na | na | na |
| | | | | |
| <u>Interaction Terms</u> | | | | |
|   Choice: Yes × Prize: Cash | | −0.46* | | −0.27 |
| | | (0.28) | | (0.43) |
|   Choice: Yes × Prize: Non-Cash | | −0.36 | | 0.85** |
| | | (0.31) | | (0.42) |
|   Choice: Yes × Prize: Indifferent | 0.11 | −0.06 | 0.32 | 0.41 |
| | (0.20) | (0.23) | (0.29) | (0.11) |
| Intrinsic Motivation | −0.00 | −0.02 | 0.07 | 0.10 |
| | (0.05) | (0.05) | (0.13) | (0.11) |
| Intercept | 2.58*** | 2.41*** | 4.33*** | 4.37*** |
| | (0.13) | (0.17) | (0.20) | (0.22) |
| Controls Included | Yes | Yes | Yes | Yes |
| Num. obs. | 208 | 208 | 208 | 208 |
| Adj. $R^2$ | 0.02 | 0.02 | | |
| AIC | | | 2597.51 | 2589.35 |
| Log Likelihood | | | −1286.76 | −1280.68 |

***$p < 0.01$; **$p < 0.05$; *$p < 0.1$

### Summary

We find a significant *main effect* showing that choice improves idea quality (confiming H1a) but not effort (no support for H1b). While the coefficient for the main effect of choice on effort is insignificant, the direction of the effect is positive. We find that choice significantly reduces the effectivness of cash incentives on quality and amplifies the positive effect of the non-cash incentive on effort.

## Study 3: Moderator of For-Profit and Non-Profit Context

In Study 3, we explore if market context *moderates* the effect of incentives on quality and effort. We conducted a crowdsourcing contest of a typical for-profit organization to generate ideas that could be commercialized to ensure future profits. To create a realistic experiment, we replicated the ideation task of the Intel Future Contest, a real-world ideation challenge sponsored by Intel. In that contest, Intel solicited ideas for new products or services around a new technology [55]. This technology aims to allow the building of a new generation of smart devices (e.g., smart wearable technology) and applications. The task provided background information on the new technology adapted from the Intel Future Contest and then we asked ideators to provide information on product features, benefits, uses, and design proposals. We advertised an open call to a public audience on Craigslist, Reddit, Facebook, and various other technology-related communities and blogs.

### Method

**Procedure.** We implemented an online experiment which randomly assigned ideators following the same general setup as in Study 2. The design is a 2 (incentive: cash, non-cash) × 2 (choice: choice, assigned) × 2 (context: for-profit, non-profit) between-subject study, resulting in six treatment conditions (Appendix 7). We manipulated market context by framing the ideation task as being solicited by either a for-profit or non-profit organization. The for-profit condition framed the ideation task as "a for-profit multinational corporation wants to develop a for-profit product or service", while the non-profit condition solicited ideas for "a non-profit research organization wants to develop a non-profit product or service."

To increase the stakes compared to Study 2, we increased the prize money to $1,000, split as follows: $500 for 1st prize, $300 for 2nd prize, and $200 for 3rd prize. We performed (streaming) random assignment of participants to treatment conditions as is common in online experiments where the total number of participants is not known ex ante. We stopped the experiment after four weeks. The study was pre-registered before data collection began (https://osf.io/8qw7t/).

**Measures.** We measured the effort in seconds spent on the ideation task. Following the same procedure as in the other studies, we collected data to measure idea quality from an outside panel recruited through Amazon Mechanical Turk. We collected 770 ratings of quality from 46 different raters who performed an average of 17 ratings each and five ratings per idea. Inter-coder reliability is excellent (0.77); Cronbach's alpha is good (0.80).

### Results

A total of 120 ideators completed the experiment and submitted at least one idea (153 ideas were submitted in total). In a post-experiment self-report question, the manipulation of the profit motive was effective with 76% of ideators correctly recalling the profit structure of the organization sponsoring the contest (i.e., for-profit vs. non-profit).

**Quantifying incentive preferences across market contexts.** Despite qualitatitive differences, we find no statistically significant difference of preference for cash or non-cash incentives between for-profit and non-profit contexts (Appendix 8, Model 2). In the non-profit context, 31% chose the non-cash incentive compared to 23% in the for-profit context. Within the for-profit context, a simple for equal proportion indicates that cash is the significantly more popular choice (cash is significantly more popular than non-cash; $p = 0.001$). There is no significant difference within the non-profit context.

**Moderating effect of market context.** We find no significant direct effect of market context on quality (Table 4: Model 1; $\beta = -.06$; *n.s.*) but a significant effect on effort (Model 6; $\beta = -.41$; $p < 0.05$). We find no significant *moderation* effect between market context and incentives for quality (Model 4; $\beta = -.21$; *n.s.*). However, the for-profit context significantly reduces the effect of non-cash incentives on effort (Model 9; $\beta = -.75$; *p < 0.05*). That is, we find evidence that market context moderates the effect of incentives on effort (H2b) but not quality (H2a).

Table 4. Study 3 – Regression analysis. Omitted category: Assigned cash, non-profit.

| Dependent Variable | Idea Quality | | | | | Effort | | | | |
|---|---|---|---|---|---|---|---|---|---|---|
| | (1) | (2) | (3) | (4) | (5) | (6) | (7) | (8) | (9) | (10) |
| Treatments | | | | | | | | | | |
| Choice: Yes | −0.18 | −0.15 | −0.18 | −0.19 | −0.15 | −0.15 | −0.12 | −0.18 | −0.14 | −0.12 |
| | (0.11) | (0.11) | (0.11) | (0.11) | (0.12) | (0.21) | (0.20) | (0.21) | (0.21) | (0.20) |
| Prize: Non-Cash | −0.33*** | −0.36*** | −0.33*** | −0.33*** | −0.35*** | −0.06 | −0.09 | −0.08 | −0.03 | −0.06 |
| | (0.12) | (0.12) | (0.12) | (0.12) | (0.12) | (0.20) | (0.20) | (0.20) | (0.20) | (0.20) |
| Context: For-Profit | −0.06 | −0.04 | −0.05 | −0.09 | −0.03 | −0.41** | −0.41** | −0.46** | −0.50*** | −0.45** |
| | (0.10) | (0.10) | (0.10) | (0.11) | (0.11) | (0.18) | (0.19) | (0.19) | (0.19) | (0.19) |
| Interaction Terms | | | | | | | | | | |
| Choice: Yes × Prize: Non-Cash | | 0.46* | | | 0.43* | | 0.41 | | | 0.37 |
| | | (0.24) | | | (0.24) | | (0.39) | | | (0.41) |
| Choice: Yes × Context: For-Profit | | | −0.05 | | −0.10 | | | 0.30 | | 0.19 |
| | | | (0.22) | | (0.23) | | | (0.39) | | (0.37) |
| Prize: Non-Cash × Context: For-Profit | | | | −0.21 | −0.25 | | | | −0.75** | −0.81** |
| | | | | (0.24) | (0.24) | | | | (0.37) | (0.39) |
| Choice: Yes × Prize: Non-Cash × Context: For-Profit | | | | | 0.37 | | | | | 0.89 |
| | | | | | (0.42) | | | | | (0.76) |
| Social Value Orientation | 0.32 | 0.25 | 0.31 | 0.30 | 0.22 | −0.23 | −0.30 | −0.15 | −0.37 | −0.41 |
| | (0.20) | (0.20) | (0.21) | (0.19) | (0.20) | (0.38) | (0.35) | (0.36) | (0.36) | (0.36) |
| Intercept | 1.65*** | 1.60*** | 1.67*** | 1.70*** | 1.78*** | 5.96*** | 5.90*** | 5.74*** | 6.18*** | 6.18*** |
| | (0.24) | (0.23) | (0.27) | (0.22) | (0.27) | (0.36) | (0.35) | (0.34) | (0.39) | (0.41) |
| Controls Included | Yes | Yes | Yes | Yes | Yes | Yes | Yes | Yes | Yes | |
| Num. obs. | 120 | 120 | 120 | 120 | 120 | 120 | 120 | 120 | 120 | 120 |
| Adj. R² | 0.13 | 0.16 | 0.13 | 0.13 | 0.14 | | | | | |
| Log Likelihood | | | | | | −951.22 | −950.58 | −950.86 | −948.80 | −947.31 |

***$p < 0.01$; **$p < 0.05$; *$p < 0.1$

Further, there is no significant effect of the three-way interaction on either idea quality (Model 5; $\beta = .37$; *n.s.*) or effort (Model 10; $\beta = .89$; *n.s.*). This suggests that the effectivness of offering a choice (such as the positive interaction effect between choice and non-cash on quality) does not strongly depend on the market context, albeit our statistical power for this analysis is quite low.

**Summary**

In Study 3 we investigate whether market context *moderates* the effect of incentives on quality (H2a) and effort (H2b). We find support for this moderation effect for effort but not quality.

## Study 4: Boundary Condition of Incentive Choice Effect

Finally, we test an important boundary condition in our last experiment (Study 4). So far, our evidence suggests that the main effect of offering ideators a choice derives from achieving improved fit between ideators' preferred incentive and the incentive they actually receive (as opposed to the direct effect of simply asking them to reveal their preference). That is, actually observing diverse incentive preferences in the population may be crucial for the main effect of incentive choice to unfold. This is important because over time some online platforms may evolve an environment in which individuals have homogeneous incentive preferences despite representing diverse populations in terms of gender, home country, and economic background. In this study, we test this boundary condition by repeating the basic setup from Study 2, using a sample of ideators drawn from a population of gig workers narrowly focused on earning income. We recruited participants from Amazon Mechanical Turk (AMT), an online labor market Here, we expect participating workers to have a homogeneous preference for cash as transactions in the online labor market are based on strict market relationships.

Prior research has shown that AMT workers are predominantly motivated by financial incentives [49]. For example, they often give themselves daily or weekly quotas of how much money they want to earn working on the AMT platform. If we find that the majority of ideators make the same choice and we find no main effect of incentive choice, this suggests that heterogeneous incentive preferences are an important boundary condition to realize a positive direct effect of incentive choice.

### Method

**Procedure**. We recruited 160 workers from the AMT online labor market. Workers were compensated with a $1 "show-up" fee and randomly assigned to the same ideation task and treatment conditions from Study 2: no prize (N=21), cash (N=20), non-cash (N=29), a choice between cash and non-cash (N=37), and a choice between cash and opt out (N=46).

### Results

**Quantifying incentive preferences.** As predicted, we find that participants drawn from AMT showed no interest in non-cash prizes whatsoever (Appendix 9). Zero participants chose the non-cash prize in the cash/non-cash condition (out of 29) and only two (out of 46) chose to opt out.

**Main effect of incentive choice on quality and effort.** We used OLS regression to explore the incentive and choice effects on idea quality and a negative binomial regression for the effects on effort. Notice that we only included the cash, assigned non-cash, and no-prize control groups in our sample, and omitted the "opt out" and "chose non-cash" groups due to the small number of individuals making those choices. We find no significant main effect of offering a choice on either quality (Table 5: Model 1; $\beta = .1$; *n.s.*) or effort (Model 2; $\beta = -.02$; *n.s.*), indicating that the mere choice of the preferred cash option – without interest in other incentive options – offered no benefit and did not lead to increased performance. To directly test if increased agency and self-determination is a plausible mechanism behind the choice effect, we added the intrinsic motivation battery to the post-experiment survey. We find no difference in intrinsic motivation between individuals who chose cash compared to those who were assigned cash ($t = .69$; d.f. = 31.14; $p = .50$), suggesting that simply asking cash motivated individuals to reveal their cash motivation does not itself lead to increased motivation.

Table 5. Study 4 – Regression analysis of AMT sample. Note that we do not estimate coefficients for the small number of ideators who chose non-cash or were indifferent as those groups are too small for a meaningful analysis. Consequently, *Choice: Yes* implies *Prize: Cash* and there is no separate interaction term that can be estimated.

| Dependent Variable | Idea Quality | Effort |
|---|---|---|
| | (1) | (2) |
| Treatments | | |
|   Choice: Yes | 0.11 | −0.01 |
| | (0.11) | (0.12) |
|   Prize: Cash | 0.07 | 0.29 |
| | (0.18) | (0.19) |
|   Prize: Non-Cash | 0.45*** | 0.32* |
| | (0.17) | (0.18) |
| Intrinsic Motivation | 0.01 | −0.01 |
| | (0.03) | (0.03) |
| Intercept | 2.93*** | 4.56*** |
| | (0.18) | (0.22) |
| Controls Included | Yes | Yes |
| Num. obs. | 175 | 175 |
| Adj. $R^2$ | 0.02 | |
| Log Likelihood | | −945.81 |

***$p < 0.01$; **$p < 0.05$; *$p < 0.1$

### Summary

We find that the population of workers on Amazon Mechanical Turk had a homogeneous preference for the cash prize. We find no sign that simply asking ideators to reveal their incentive preference increased their motivation. Together, the findings suggest that the benefit of offering a choice unfolds through a better incentive-preference fit. The increased sense of autonomy from the choice itself does not lead to higher performance.

## General Discussion and Theoretical Contribution

The research presented in this paper sets out incentive choice (cash vs. non-cash) to alleviate a fundamental issue in incentive design and enhance the alignment between incentive and preferences to increase effort and idea quality in crowdsourcing contests across market contexts (for-profit vs. non-profit).

    We present evidence from four consecutive empirical studies (see result summary in Table 6). Our field study (Study 1) helped us *quantify* ideators' preferences and suggests they are very diverse, with over 49% choosing among one of the non-cash incentives and 28% choosing cash. There may be several explanations for the no-choice effect: 1) participants who did not make a choice were fine with the default cash option and did not want to explicitly reveal their incentive preference; 2) participants may have been indifferent and had no strong preference for any of the offered incentives; 3) they made no choice because none of the offered incentives matched their preferences; 4) they were opposed to rewards and/or choice altogether and were happy to participate without any incentive.

Table 6. Summary Findings.

| | Key Findings |
|---|---|
| **Study 1** (Field study n=1,205) | • **Quantify** incentive preferences: > 49% choosing among one of the non-cash incentives; 28% choosing cash; 23% prefer to make no choice at all.<br>• Preliminary evidence of main effect of choice on quality |
| **Study 2** (Online experiment with random assignment n=208) | • Choice increases quality (**main effect**) but not effort (supporting H1a)<br>• Choice reduces the effect of cash incentives on quality and increases the effect of non-cash incentives on effort. |
| **Study 3** (Online experiment with market context as treatment manipulation, n=120) | • Market context **moderates** the effect of incentives on effort (but not quality; supporting H2b)<br>• No significant difference in preference for non-cash prize in for-profit context. Lower effort for non-cash incentive in for-profit context. |
| **Study 4** (Online experiment with gig-worker sample, n=160) | • Establishes heterogeneous preferences as important **boundary condition**: no interest in non-cash incentives in pool of gig-workers focused on earning income.<br>• Without sorting, no effect materializes. Indicating improved incentive-preference fit is driving mechanism. |

The field study (Study 1) and the randomized lab experiment (Study 2) both provide evidence for the *main effect* of offering a choice on idea quality (not effort; both those studies were set in a non-profit context). Study 3 finds that market context *moderates* the effect of incentives on effort (but not idea quality). One possible explanation for the null-effect on quality may be that idea quality in creative settings only partially depends on effort or it may simply be a result of low statistical power (while not statistically significant, the regression coefficient for quality points in the same direction as that for effort). Ideators exerted less effort in for-profit settings than in non-profit settings overall and even less when the for-profit setting is paired with a non-cash incentive. Finally, we point to an important boundary condition (Study 4): If ideators have uniform incentive preferences for cash such as gig workers on Amazon Mechanical Turk, offering a choice of incentives has no effect. The lack of a direct effect of choice per se, suggests that the benefit arises from the improved matching between incentive preference and the incentive being offered. As a result, we theorize that the gains from offering a choice do not arise from a feeling of agency but instead from improved sorting of preferences to incentives. Without heterogeneous incentive preferences, there is no room for gains from sorting of preferences to actual incentive. This suggests that the strength of the effect of offering a choice depends both on the diversity of ideators' incentive preferences and the diversity of incentives being offered to maximize this sorting effect. Across our four studies, personal characteristics such as gender, economic background, and intrinsic motivation served only as weak proxies for incentive preferences while social value orientation emerged as a strong predictor in Study 3.

Our findings make three main contributions to theory. First, past work on crowdsourcing design focused on contest design aspects like prize structures [46,54,64], number of contestants [11], and entry barriers [26]. By contrast, we establish that incentive choice is a pivotal aspect of incentive design that is little understood [53] and are the first to shed light on the underlying mechanism why it may work. Our study explains why offering ideators a choice of incentives per se can improve performance. We theorize that offering a choice improves creative performance in crowdsourcing contests because it improves the incentive-preference fit which increases effort and performance rather than an increased sense of autonomy and sense of control [15]. This connects with the idea of incentive choice in research on mass customization, where customers self-configure products that match their preferences rather than choose standardized products [22]. Our study is the first to shed light on the mechanism behind the effectiveness of incentive choice and its important boundary conditions. Outside crowdsourcing, [14,57] are notable exceptions that explored incentive choice in a lab experiment giving participants a choice between fixed and performance-based pay.

We contribute to work on crowdsourcing design by explicating an important boundary condition: offering a choice is ineffective when the incentive preferences in the target population are homogeneous. Incentive preferences may be homogeneous despite diverse geographic and economic background when recruiting ideators from online labor markets such as Amazon Mechanical Turk (see Study 4). Further, the sorting effect seems to strengthen when many attractive incentive options are offered (Study 1; c.f. [68]). Additionally, ideators may enjoy various benefits [9] and may hence also be rather indifferent to the available incentive options. Thus, forcing participants to reveal their preferences by choosing their preferred incentive may only increase their burden and not offer additional value [50]. Incentive choice may further consider an opt-out option (see Study 2) as participants may prefer to forgo any incentive rather than accept an incentive that, in their eyes, does not match their demonstrated performance [21].

Second, our study contributes to a better understanding of incentives in different market contexts. While crowdsourcing contests are used equally in for-profit vs. non-profit market contexts, existing research has studied the direct effect of cash vs. non-cash incentives [6,24], but not considered market context as a moderator. Our study fills this gap and extends received knowledge that incentives can sometimes backfire when they are misaligned with the market context in which they are used [29,65]. We show that incentive preference and its effect on quality and effort is not only influenced by individual's motives and personal characteristics, but also market context. Researchers [34] have applied Fiske's relationship theory to explain the signaling effect of incentives and their influence on effort [63], and empirical studies have referred to different effects of incentives in for-profit and non-profit contexts [12,16]. However, no one has yet considered the classification of context as a monetary market (for-profit) vs. social market (non-profit) as an important moderator for predicting incentive preference and its effect on quality and effort.

Third, we expand on previous research that has identified a variety of intrinsic and extrinsic reasons for why individuals engage in crowdsourcing contests [1,8,11,23,45] by quantifying of the extent to which this this occurs and demonstrating that individuals not only have diverse preferences but actually choose different incentives when given the choice. While our results are consistent with past research showing that cash is generally the most prevalent single incentive due to its high option value [36], in some settings almost half prefer non-cash incentives. Our work is one of the first to validate various incentive preferences in a field setting. This emphasizes the importance of heterogeneous incentives not only as a niche aspect but a core driver of motivation to participate in crowdsourcing contests. This insight opens opportunities to improve contest design by considering alternative incentives. We also determine that ideators not only have diverse incentive preferences but sometimes even prefer to opt out of receiving any incentives and sometimes are indifferent, thus suggesting entirely new forms of incentives to consider. These findings emphasize the existing literature that it is not easy to offer an appropriate incentive upfront [48,56] and providing incentive choice may be useful.

## Managerial Implications

Our research has four practical implications for open innovation managers to design more effective incentive regimes. First, offering a choice of incentives may increase the effectiveness of incentives in crowdsourcing contests, especially when incentive preferences are diverse and a set of suitable non-cash incentives are available. Offering a choice can alleviate the concern that managers may not know what the most desirable incentive is and may worry about missing out if cash is not offered. Second, the incentive choice should be implemented as an optional choice with a clearly defined cash default option to cater to ideators who may be indifferent. This choice allows contest organizers to offer unique and unexpected incentives that may be very effective in special contexts (e.g., NASA offering a low value artifact like a sticker mentioning "flown in space"; [9,63]). Third, the accentuation of a non-profit context (social market) matters. Social markets can positively affect participants' level of effort. Thus, crowdsourcing contests in social markets should underline their social character. Fourth, heterogeneous incentive preferences may over time vanish as online platforms specialize and evolve to cater to a more homogeneous user group (e.g., AMT). Crowdsourcing platforms must be aware that if they heavily rely on cash as their standard incentive, it will not be surprising that their community expects cash. Conversely, offering non-cash incentives and hosting contests for a variety of market contexts can be a means to attract heterogeneous ideators, which may improve idea quality throughout the platform.

## Limitations and Future Research

Our findings are not without limitations. Although gathered in various setups and under realistic conditions, further research is required to establish the dimensions in which they generalize. Our strongest findings in favor of non-cash incentives come from the field data in Study 1. This study was set in a non-profit context and those results may not fully generalize to for-profit settings despite our insights from Studies 2-4. Future research could evaluate if similar incentive preferences persist in crowdsourcing contests in for-profit contexts. While our results are consistent in regards to effect of incentives, choice, and context, more research is required to investigate the effect in additional settings and to enhance the overall applicability of our conclusions. In addition, numerous new areas necessitate further investigation, including design of choice options, the effect of indifference, the offering of incentive bundles, the effect of incentive opt out, and the conditions that create a social market character.

Further, exploring how incentives over time lead to homogeneous preferences and adjusted behaviors, e.g., those found at AMT, would be illuminating. Additionally, our study employed prize purses that are commonly used in current reserach. Higher prizes may lead to different outcomes and different incentive choices. In particular, we expect incentives to function differently in crowdsourcing contests compared with grand challenges like NASA's $1M CO2 Conversion Challenge or the $10M Ansari X-Prize for Suborbital Flight. We speculate that ideators would be much less likely to either forgo incentives or choose a non-cash option if the stakes are very high. Consequently, the full range of ideators' sensitivity to prize levels remains an open question. Our analysis also focused on shifts in the mean quality of ideas due to selection and treatment effects. However, sometimes shifts in maximum quality are more important than shifts in mean quality, especially in rank-order.

## Conclusion

In conclusion, this research significantly advances our understanding of incentive design in crowdsourcing contests, highlighting the importance of offering a choice between cash and non-cash incentives to match diverse ideators preferences. It underscores the role of market context which moderates the effectiveness of these incentives and reveals that personal characteristics are less indicative of preference than previously thought. The findings open new pathways for designing more effective crowdsourcing contests, emphasizing the need for flexibility and customization of incentives to cater to diverse participant motives and market contexts.

# Online Appendices

Appendix 1. Positioning of Current Paper in Literature on Choice of Incentives

| Dependent Variable | Selected References | Incentive | | | Application | Study Design | Context | Summary |
|---|---|---|---|---|---|---|---|---|
| | | Choice of | Intrinsic (non-cash) | Extrinsic (cash) | Creativity in Crowdsourcing | Observational and Experimental Data | For profit & Non-profit (both) | |
| Creative & Innovation Performance | Current article | Yes | Yes | Yes | Yes | Yes | Yes | The current study tests the effectiveness of cash vs. non-cash incentives – assigned to or self-selected by ideators – in crowdsourcing contest for-profit vs. non-profit organizers. |
| | [26] | Yes | Yes | Yes | Yes | Yes | No | The authors test in a single experiment with two rounds how providing participants with incentive choice (monetary vs. symbolic) choice impacts solution quality. |
| | [6] | No | Yes | Yes | Yes | No | No | Cappa et al. empirically test if two different types of rewards – monetary and social rewards – increase the number of contributions in crowdsourcing. |
| | [12] | No | Yes | Yes | No | No | No | Eisenberger and Rhoades examined unrewarded and rewarded creativity training and its impact on creative task performance. |
| | [20] | No | Yes | Yes | No | Yes | No | Heymann and Ariely test the relationship between forms of compensation (cash vs. token), the levels of payment (no, low, and medium), and the resulting effort expended in monetary and social markets. |
| | [31] | No | Yes | Yes | No | No | No | Sittenthaler and Mohen employs an experiment to test the impact of monetary, non-monetary, and a combination of monetary and non-monetary incentives on performance. |
| | [5] | No | No | Yes | Yes | Yes | No | The authors test incentives and their impact on contest performance in high and low-uncertainty problems with greater and lower rivalry. |
| | [21] | No | No | Yes | Yes | Yes | No | The authors empirically test if incentive and parallel path effects – adding numbers of competitors – are of comparable magnitude and thus be explicitly considered together when designing crowdsourcing contests. |
| | [33] | No | No | Yes | Yes | Yes | No | Toubia examines if tailored ideation incentives improve creative output. |
| | [1] | No | No | Yes | Yes | Yes | No | Acar investigates whether the use of monetary rewards is effective in stimulating creativity and, if so, how large those rewards should be. |
| | [25] | No | No | Yes | Yes | No | No | The authors explore the effects of different incentives on crowdsourcing participation and contribution quality in randomized field experiments. |
| | [28] | No | No | Yes | No | Yes | No | The authors empirically test the effects of extrinsic financial rewards on intrinsic motivation. |

| Dependent Variable | Selected References | Incentive | | | Application | Study Design | Context | Summary |
|---|---|---|---|---|---|---|---|---|
| | | Choice of | Intrinsic (non-cash) | Extrinsic (cash) | Creativity in Crowdsourcing | Observational and Experimental Data | For profit & Non-profit (both) | |
| Creative & Innovation Performance | [8] | No | No | Yes | No | Yes | No | Deci tests in two laboratory experiments and one field experiment the intrinsic motivation to perform an activity. |
| | [15] | No | No | Yes | No | No | No | The authors investigate synergistic extrinsic motivators to foster creativity intrinsically motivated knowledge workers. |
| | [14] | No | No | Yes | No | No | No | Erat and Gneezy empirically test whether piece-rate and competitive incentives so, how the incentive effect depends on different types – and not merely incentives. |
| | [13] | No | No | Yes | No | No | No | Eisenberger and Selbst investigate why behaviorists and cognitive oriented investigators show opposite conclusions about reward's effects on creativity. |
| | [3] | No | No | Yes | No | No | No | The authors examine the effect of reward on children's and adults' creativity. |
| | [2] | No | No | Yes | No | No | No | Amabile test the creativity motivation hypothesis by investigating the effects of a common extrinsic constraint – competing for prizes – on children's artistic creativity in a field setting. |
| | [27] | No | No | Yes | No | No | No | Pinder test additivity versus non-additivity of intrinsic and extrinsic incentives on work motivation, performance and attitude. |
| | [9] | No | No | Yes | No | No | No | Deci empirically investigates what happens to a person's intrinsic motivation for an activity when he is rewarded extrinsically for performing the activity. |
| | [19] | No | No | Yes | No | No | No | Hammer and Foster test that contingent monetary rewards actually reduced intrinsic task motivation in both a boring and nonboring task setting. |
| | [11] | No | No | Yes | No | No | No | Eder and Manso test in a controlled experimental setting the effects of different incentives schemes (e.g. fixed wage, pay for performance, exploration) on innovation and performance. |
| | [16] | No | No | Yes | No | No | No | The authors evaluate in a series of laboratory experiments the fixed prize mechanisms as a means to obtain a given quality of research at as low a cost as possible under various market conditions. |
| | [10] | No | No | No | No | No | No | Deci and Cascio test changes in intrinsic motivation as a function of negative feedback and threats. |

Our review categorizes the studies along incentives and choice. In addition, we classify the study settings (creativity in crowdsourcing vs. other settings), how the data was gathered (observational vs. experimental data design), and the context setting (for-profit vs. non-profit). While our review does not claim to be exhaustive it covers the most relevant and actual studies.

Appendix 2. Study 1

*Estimation Procedure:* We estimate the following system of three simultaneous latent equations

$$y_i^{S*} = \beta^{S\prime} x_i^S + \epsilon_i^S \qquad (1)$$
$$y_i^{O1*} = \beta^{O1\prime} x_i^{O1} + \epsilon_i^{O1} \qquad (2)$$
$$y_i^{O2*} = \beta^{O2\prime} x_i^{O2} + \epsilon_i^{O2}. \qquad (3)$$

Equation (1) is the *selection* rule, where an individual's $i$ choice $y_i^S$ is the choice of the cash incentive or the non-cash incentive. $y^{O1*}$ and $y^{O2*}$ are the latent outcomes, only one of which is observable, depending on the sign of $y_i^S$ (that is: we observe either the quality of designs produced under the cash treatment or the quality produced under the non-cash treatment but not both). Hence, we observe

$$y_i^S \begin{cases} 0 & \text{if } y_i^{S*} < 0 \\ 1 & \text{otherwise} \end{cases} \qquad (4)$$

$$y_i^O \begin{cases} y_i^{O1*} & \text{if } y_i^S = 0 \\ y_i^{O2*} & \text{otherwise.} \end{cases} \qquad (5)$$

Covariates $x_i^S$ are fixed ideator characteristics (home-country GDP, gender, and western background) and a measure of competition at the time of registration (number of designs that had already been submitted). The vector $\beta^{S\prime}$ are the estimated regression coefficients and $\epsilon_i^S$ are the error terms for the selection equation. Equation (2) and (3) are the *outcome* equations and model design quality conditional on covariates $x_i^O$ (professional designers, time since registration, and counts of ideators' submitted designs, comments, and ratings). The outcome equations are estimated separately for each treatment: Eq. (2) is estimated for ideators who choose a non-cash prize with the observed dependent variable $y_i^{O1}$ and Eq (3) is estimated for ideators who choose the cash prize with the observed dependent variable $y_i^{O2}$. The two vectors $\beta^{O1\prime}$ and $\beta^{O2\prime}$ are separate sets of estimated regression coefficients for the covariates $x_i^{O1}$ and $x_i^{O2}$, respectively (note, however, that we use the same covariates in both; that is $x_i^{O1} = x_i^{O2}$). The error terms $\epsilon_i^S$, $\epsilon_i^{O1}$, and $\epsilon_i^{O2}$ are trivariate normally distributed with 0 mean and covariances [32] given by

$$\begin{bmatrix} \epsilon^S \\ \epsilon^{O1} \\ \epsilon^{O2} \end{bmatrix} \sim N(0, \Sigma), \tag{6}$$

with $\Sigma$

$$\Sigma = \begin{bmatrix} 1 & \rho_1 & \rho_2 \\ \rho_1 & 1 & \rho_{23} \\ \rho_2 & \rho_{23} & 1 \end{bmatrix}. \tag{7}$$

In the model, the presence of selection can be quantified by the statistical and substantive significance correlation coefficient $\rho_1$ between the errors of the selection equation (Eq. 1), and the outcome equations for ideators who choose a non-cash prize (Eq. 2) and the correlation coefficient $\rho_2$ between the errors of the selection equation (Eq. 1), and the outcome equations for ideators who choose the cash prize (Eq. 3). If $\rho_{1/2}$ is zero, then the unmeasured factors which influence whether an ideator chooses an incentive are independent of the unmeasured factors which determine the quality of the design produced by that ideator. If $\rho_{1/2}$ is positive then the unmeasured factors that lead an ideator to choose an incentive are positively correlated with the unmeasured factors that lead them to produce designs of higher quality. If, on the other hand, $\rho_{1/2}$ is negative then the unmeasured factors that lead an ideator to choose an incentive are negatively correlated with the unmeasured factors that lead them produce designs of higher quality. We estimate the equations simultaneously using maximum-likelihood in *R* [29] using the *SampleSelection* package [32].

*Construction of Dependent Variable:* In this section we provide additional details how we constructed the baseline rating of design quality. We measure the dependent variable, Design Quality, for all designs submitted to the contest using the Consensual Assessment Technique (CAT; Amabile, 1982). We recruited an independent jury that was blind to the research hypotheses from reliable and experienced workers on Amazon Mechanical Turk.[1]

This panel evaluated each product based on the following six dimensions: (1) creativity, (2) novel use of materials (e.g., materials are used in a unique way), (3) novel association (e.g., unique or unusual association with existing products or objects), (4) variation of materials used (e.g., different materials, number of colors, originality), (5) level of detail and complexity (e.g., of the design or decoration), and (6) appearance (i.e., how good it would look in a home or office).

Workers were instructed to make relative assessments based on their own definition of creativity [2]. As part of the instructions, workers were shown a grid of nine randomly selected designs to facilitate this relative assessment. Workers evaluated designs in random order and for each design, the assessment items were arranged in random order. We collected five evaluations for each design, resulting in 2,927 ratings of each of the six assessment items (17,562 ratings in total) from a total of 77 different raters. We apply the technique suggested by (Ipeirotis, Provost, and Wang 2010) to identify and then remove unreliable workers. The technique identified 17 low-quality raters who submitted ratings with extremely low information content (e.g., rating completely at random or submitting

---

[1] AMT offers a mechanism to restrict the pool of eligible workers using various qualifications. We restricted our task to workers with the following qualifications:
- Approval rate for all prior tasks greater than or equal to 99%
- Number of tasks approved greater than or equal to 10,000.

That is, selecting experienced workers is nothing that we did specifically, but is a feature directly available on AMT.

identical ratings for all five items). These raters collectively submitted 432 ratings (15%). After cleaning the ratings from low-quality raters, 2,495 ratings remain (that is, 14,970 ratings in total), with an average of 4.3 ratings per design and 42 ratings per rater.2 We paid workers on average $4.20 for their effort.

The key premise of a crowdsourcing contest is to attract submissions from a diverse pool of participants [23]. Hence, crowdsourcing contests are most effective when they are geared toward reaching out to outsiders to solicit creative ideas. Research has now shown that 1) online panels such as those from Amazon Mechanical Turk are appropriate [34], 2) expertise does not significantly affect the quality of assessments [30], 3) high correlation exists between the assessments from experts and laypeople across different evaluation methods [4], and 4) assessments from panels of consumers can be even better than expert panels [24]. Overall, we conceptualize the performance of ideators in crowdsourcing idea contests as the quality of the best idea that an individual submits, rather than effort (e.g., measure in time spent on the task or the length of the submitted idea).

Prior research on ideation has often defined performance as the average quality of ideas or the number of ideas generated by an individual, ignoring that most organizations seek a few great ideas [17]. Consequently, in case an ideator made multiple design submissions, we use the quality of the ideator's best submission. That is, at the individual level, the performance of an ideator is then measured as the quality of the best idea that an individual submitted. A focus on ideators' best idea rather than average idea quality is also more consistent with the nature of a rank-order tournament in which prizes are only awarded to the

---

2 Robustness tests including all ratings and not dropping low-quality raters do not substantively change our conclusions but explained variance ($R^2$) is lower, supporting the notion that low-quality raters added only noise.

contest winners.[3]

Cronbach's alpha of 0.9 indicates high internal consistency of the six assessment items. Intercoder reliability ICC (2,k) for the aggregated scale (all six items) is 0.70 indicating good inter-rater agreement [7]. Individual item ICC (2,k) range from 0.59 to 0.66. The quality of ideas, i.e., their creativity, is critically important for an innovation's success and ultimately market success [24] and is thus most important from a managerial perspective.

---

[3] We do perform robustness tests using an ideator's average quality instead and find substantively similar results. See section on robustness tests for more details.

*Sample Designs* – *We s*how example designs submitted to the consumer innovation contest.

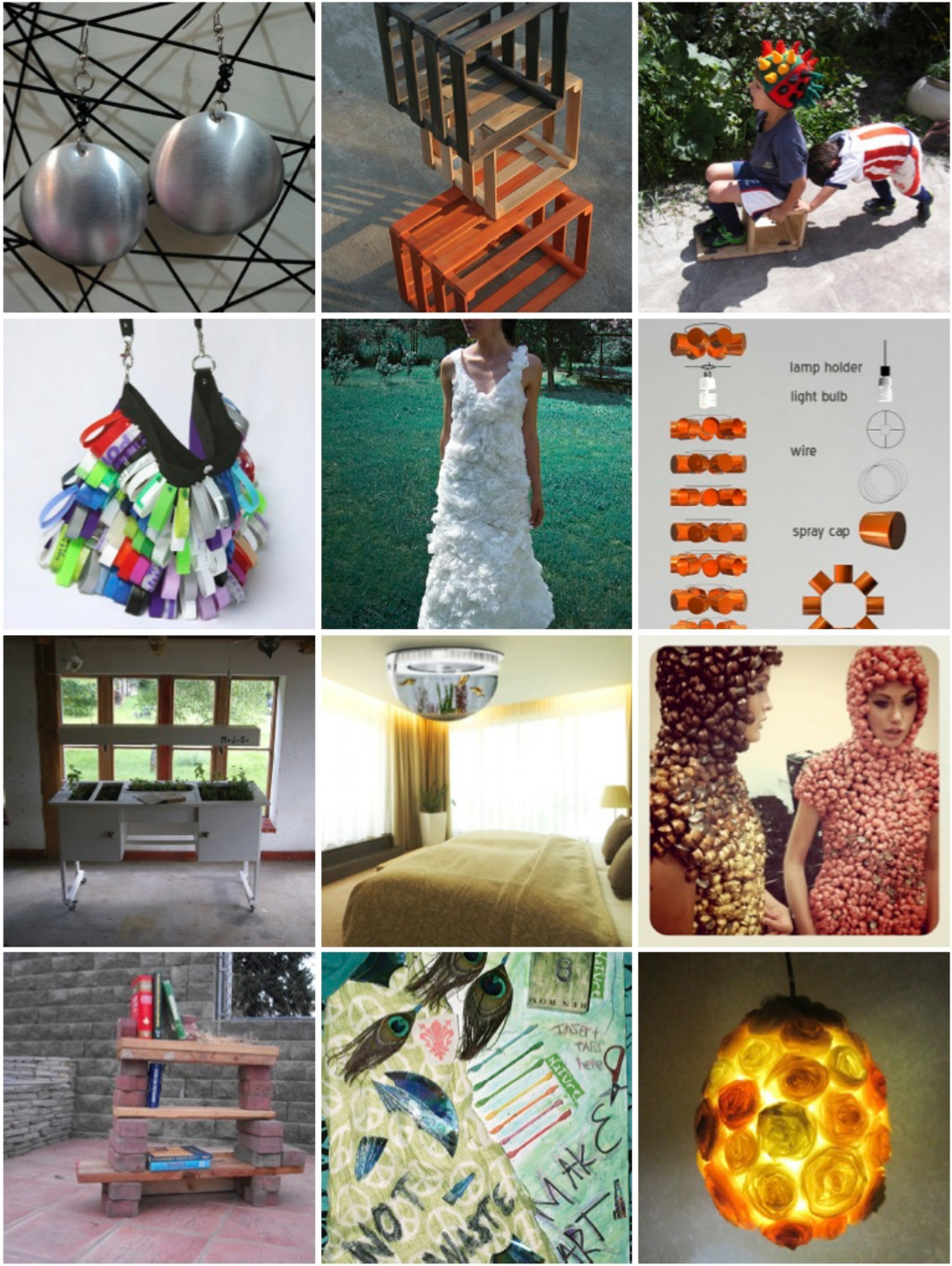

*Robustness Tests Using Alternative Measures of Quality:* Since the goal in rank order contests is to win the contest, we focused the analyses presented in the main paper on the quality of the best idea submitted by an ideator. We show correlation coefficients of three different quality measures in the table below and find substantively similar results for any of the three quality measures. Not surprisingly, the maximum and mean quality are highly correlated ($\rho = 0.88$; $p < .001$). We find substantively similar results using average quality as the dependent variable.

Correlation of quality measures (N=259).

|  | Mean | SD | Min | Max | (1) | (2) |
|---|---|---|---|---|---|---|
| Design Quality Best Idea (AMT; main measure used in paper) (1) | 3.33 | 0.66 | 1.00 | 4.78 | | |
| Average Design Quality (AMT) (2) | 3.14 | 0.61 | 1.00 | 4.50 | 0.88*** | |
| Average Community Rating Best Idea (3) | 3.56 | 0.73 | 1.00 | 4.90 | 0.37*** | 0.28*** |

Appendix 3. Study 1 – Descriptive statistics and correlations of main study variables of ideators who made at least one design submission (N = 259).

| | Mean | SD | Min | Max | (1) | (2) | (3) | (4) | (5) | (6) | (7) | (8) | (9) | (10) | (11) |
|---|---|---|---|---|---|---|---|---|---|---|---|---|---|---|---|
| Design Quality (1) | 3.33 | 0.66 | 1.00 | 4.78 | | | | | | | | | | | |
| Designs Submitted (2) | 2.26 | 2.65 | 1.00 | 24.00 | 0.29 | | | | | | | | | | |
| Incentive: No Answer (3) | 0.55 | 0.50 | 0.00 | 1.00 | 0.08 | 0.22 | | | | | | | | | |
| Incentive: Cash (4) | 0.27 | 0.44 | 0.00 | 1.00 | -0.06 | -0.14 | -0.66 | | | | | | | | |
| Comments Written (5) | 4.46 | 10.32 | 0.00 | 115.00 | 0.14 | 0.30 | 0.21 | -0.16 | | | | | | | |
| Ratings Submitted (6) | 7.11 | 18.82 | 0.00 | 165.00 | 0.00 | 0.10 | 0.11 | -0.08 | 0.57 | | | | | | |
| Tenure (7) | 31.14 | 22.06 | 0.14 | 70.44 | 0.00 | 0.21 | 0.11 | -0.07 | 0.27 | 0.22 | | | | | |
| log(GDP) (8) | 10.28 | 0.61 | 8.44 | 11.25 | 0.02 | -0.02 | -0.07 | 0.03 | -0.09 | 0.00 | -0.06 | | | | |
| Western (9) | 0.47 | 0.50 | 0.00 | 1.00 | -0.07 | -0.06 | 0.04 | -0.04 | 0.02 | -0.07 | 0.01 | 0.40 | | | |
| Female (10) | 0.61 | 0.49 | 0.00 | 1.00 | -0.08 | 0.09 | -0.04 | 0.03 | 0.12 | -0.01 | 0.04 | 0.02 | -0.03 | | |
| Designs Prior to Registration (11) | 290.89 | 171.55 | 0.00 | 577.00 | -0.02 | -0.22 | -0.11 | 0.07 | -0.28 | -0.21 | -0.99 | 0.08 | -0.01 | -0.04 | |
| Professional (12) | 0.36 | 0.48 | 0.00 | 1.00 | -0.04 | -0.03 | -0.06 | 0.04 | -0.01 | -0.01 | 0.07 | -0.12 | 0.10 | -0.10 | -0.08 |

Appendix 2. Study 1 - Bivariate Probit of joint decision to answer the incentive question and participate (make at least one design submission). Sample: All registered ideators.

|  | Choose Incentive (1) | > 0 Effort (2) |
|---|---|---|
| Intercept | −3.208*** | 4.968*** |
|  | (0.834) | (0.918) |
| log(GDP) | 0.312*** | −0.539*** |
|  | (0.081) | (0.086) |
| Western | 0.606*** | −0.281** |
|  | (0.093) | (0.127) |
| Female | 0.514*** | −0.428*** |
|  | (0.092) | (0.094) |
| Designs Prior to Registration | −0.049 | 0.038 |
|  | (0.038) | (0.036) |
| Choose Incentive: Yes |  | 0.621** |
|  |  | (0.261) |
| $\rho$ |  | -0.863 |
| (p-value, $\chi^2$ test of $\rho = 0$) |  | (0.002)*** |
| Total edf |  | 12 |
| Num. obs. |  | 1,205 |
| Pseudo-$R^2$ |  | 0.088 |

***$p < 0.01$, **$p < 0.05$, *$p < 0.1$

**Choice and becoming active.** Interestingly, the 77% (N=924) of participants who actively chose an incentive show a lower probability to submit an idea 12.8% (N=118) than those 23% (N=281) participants who made no choice. No-choice participants show a much higher probability to submit an idea (50%, N=140; Table 1 in the main text).

To further explore the effect of incentive choice on becoming active, we conducted a bivariate probit model [18]. This model allowed us to further explore the effect of the sequential choice of incentive followed by the choice to exert effort and make a submission. The analysis shows that choice has a positive effect on becoming active when controlling for participants' individual characteristics. Personal characteristics – high-income country (β = -.54; p < 0.01), western background (β = -.28; p < 0.05), and female (β = - .43; p < 0.01) – are strongly negatively correlated with submitting a design. Controlling for these characteristics reveals a positive effect of choosing an incentive (β = .62; p < 0.05) and becoming active. Contrary to our descriptive observation, we find that choice significantly increases the likelihood of becoming active by 38.7%.

Appendix 3. Study 1 - Logistic regression of non-cash incentive preference. Sample: All registered ideators who expressed incentive preference.

| Dependent Variable | Non-Cash Preference |
|---|---|
| | (1) |
| Intercept | −0.91 |
| | (1.67) |
| log(GDP) | 0.05 |
| | (0.16) |
| Western | 0.97*** |
| | (0.18) |
| Female | 0.38* |
| | (0.19) |
| AIC | 1152.77 |
| Log Likelihood | −572.39 |
| Deviance | 1144.77 |
| Num. obs. | 924 |

***$p < 0.01$; **$p < 0.05$; *$p < 0.1$

Appendix 4. Study 2 (Volunteer Science Sample): Study design and description of participant choices.

| | Total | No Prize | Incentive Received Cash | Non-Cash | Indifferent | Comment |
|---|---|---|---|---|---|---|
| Assigned No Prize | 13 | 13 (100%) | | | | |
| Assigned Cash | 23 | | 23 (100%) | | | |
| Assigned non-Cash | 26 | | | 26 (100%) | | |
| Choice 1 (cash/non-cash) | 41 | | 23 (56%) | 18 (44%) | | no sig. difference |
| Choice 2 (cash/indifferent/non-cash) | 60 | | 22 (37%) | 16 (27%) | 22 (37%) | no sig. difference |
| Choice 3 (cash/opt-out) | 45 | 13 (29%) | 32 (71%) | | | cash sig. more popular ($p < 0.007$) |
| Total | 208 | 26 | 100 | 60 | 22 | |

Appendix 5. Study 3 (Framing Study): Study design and description of participant choices. No significant difference between choosing cash in for-profit vs. non-profit framing (30 out of 39 vs. 27 out of 39; p = 0.6.

| | Treatment Condition | Total | Incentive Received Cash | Incentive Received Non-Cash | Comment |
|---|---|---|---|---|---|
| For-Profit | Assigned Cash | 9 | 9 (100%) | | |
| | Assigned Non-Cash | 13 | | 13 (100%) | |
| | Choice (cash/non-cash) | 39 | 30 (77%) | 9 (23%) | cash sig. more popular than non-cash ($p = 0.001$) |
| Non-Profit | Assigned Cash | 9 | 9 (100%) | | |
| | Assigned Non-Cash | 11 | | 11 (100%) | |
| | Choice (cash/non-cash) | 39 | 27 (69%) | 12 (31%) | cash sig. more popular than non-cash ($p = 0.02$) |
| | | | | | cash equally popular in for-profit vs. non-profit ($p = 0.79$) |
| | Total | 120 | 75 | 45 | |

Appendix 6. Study 3: Social value orientation is strong predictor of preference for non-cash incentives across both for- and non-profit contexts.

| Dependent Variable | Non-Cash Preference | | |
|---|---|---|---|
| | (1) | (2) | (3) |
| Intercept | −3.68*** | −3.78*** | −4.19** |
| | (1.09) | (1.10) | (1.78) |
| SVO | 4.20*** | 4.13*** | 4.74* |
| | (1.56) | (1.55) | (2.62) |
| Context: Non-Profit | | 0.29 | 0.95 |
| | | (0.55) | (2.24) |
| SVO × Context: Non-Profit | | | −0.99 |
| | | | (3.25) |
| AIC | 84.55 | 86.28 | 88.18 |
| Log Likelihood | −40.28 | −40.14 | −40.09 |
| Deviance | 80.55 | 80.28 | 80.18 |
| Num. obs. | 78 | 78 | 78 |

***$p < 0.01$; **$p < 0.05$; *$p < 0.1$

Appendix 7. Study 4 (Online Labor Market Sample): Study design and description of participant choices.

| | | Incentive Received | | | | |
|---|---|---|---|---|---|---|
| Treatment Condition | Total | No Prize | Cash | Non-Cash | Indifferent | Comment |
| Assigned No Prize | 21 | 21 (100%) | | | | |
| Assigned Cash | 20 | | 20 (100%) | | | |
| Assigned non-Cash | 29 | | | 29 (100%) | | |
| Choice 1 (cash/non-cash) | 22 | | 22 (100%) | 0 (0%) | | cash sig. more popular ($p < 0.001$) |
| Choice 2 (cash/indifferent/non-cash) | 46 | | 37 (80%) | 4 (9%) | 5 (10%) | cash sig. more popular ($p < 0.001$) |
| Choice 3 (cash/opt-out) | 48 | 2 (4%) | 46 (96%) | | | cash sig. more popular ($p < 0.001$) |
| Total | 186 | 23 | 125 | 33 | 5 | |

# REFERENCES - Appendix